\documentclass[]{raa}           
\usepackage{graphicx,times}
\usepackage{natbib}
\usepackage{amssymb,amsmath}
\usepackage{booktabs}
    \usepackage{multirow} 
\bibpunct{(}{)}{;}{a}{}{,}

\usepackage[pagebackref=true]{hyperref}
\begin{document}

   \title{Forecasting Constraint on the $f(R)$ Theory with the CSST SN Ia and BAO Surveys}

 \volnopage{ {\bf 20XX} Vol.\ {\bf X} No. {\bf XX}, 000--000}
   \setcounter{page}{1}

   \author{Jun-Hui Yan 
   \inst{1,2}, Yan Gong\inst{*1,2,3}, Minglin Wang\inst{1,2}, Haitao Miao\inst{1,2}, Xuelei Chen\inst{1,2,4,5}
   }

   \institute{ National Astronomical Observatories, Chinese Academy of Sciences, Beijing 100101, People’s Republic of China; {\it gongyan@bao.ac.cn}\\
        \and
             University of Chinese Academy of Sciences, Beijing 100049, People’s Republic of China\\
	\and
Science Center for China Space Station Telescope, National Astronomical Observatories, Chinese Academy of Sciences, 20A Datun Road, Beijing 100101, China\\
\and
Department of Physics, College of Sciences, Northeastern University, Shenyang 110819, China\\
\and
Centre for High Energy Physics, Peking University, Beijing 100871, People’s Republic of China\\
\vs \no
   {\small Received 20XX Month Day; accepted 20XX Month Day}
}

\abstract{The $f(R)$ modified gravity theory can explain the accelerating expansion of the late Universe without introducing dark energy. In this study, we predict the constraint strength on the $f(R)$ theory using the mock data generated from the China Space Station Telescope (CSST) Ultra-Deep Field (UDF) Type Ia supernova (SN Ia) survey and wide-field slitless spectroscopic baryon acoustic oscillation (BAO) survey. We explore three popular $f(R)$ models and introduce a parameter $b$ to characterize the deviation of the f(R) theory from the $\Lambda$CDM theory. The Markov Chain Monte Carlo (MCMC) method is employed to constrain the parameters in the $f(R)$ models, and the nuisance parameters and systematical uncertainties are also considered in the model fitting process. Besides, we also perform model comparisons between the $f(R)$ models and the $\Lambda$CDM model. We find that the constraint accuracy using the CSST SN Ia+BAO dataset alone is comparable to or even better than the result given by the combination of the current relevant observations, and the CSST SN Ia+BAO survey can distinguish the $f(R)$ models from the $\Lambda$CDM model. This indicates that the CSST SN Ia and BAO surveys can effectively constrain and test the $f(R)$ theory.
}

   \authorrunning{J.-H. Yan et al. }            
   \titlerunning{Constraint on $f(R)$ Theory by CSST}  
   \maketitle

%
\section{Introduction}           
\label{sect:intro}

The late-time acceleration of the Universe, first observed by the Supernova Search Team \citep{SupernovaSearchTeam:1998fmf} and the Supernova Cosmology Project \citep{SupernovaCosmologyProject:1998vns}, has posed a significant puzzle in modern cosmology. General relativity (GR) is widely accepted as the fundamental theory describing the geometric properties of spacetime, with the Einstein field equations yielding the Friedman equations that describe the evolution of the Universe within the framework of general relativity. Introducing a new dark energy component in this framework has proven effective in describing standard cosmology based on radiation and matter-dominated epochs, corresponding to the conventional Big Bang model.

Theoretical efforts to account for this phenomenon within the confines of general relativity face challenges, prompting the need for novel explanations or modifications to the existing framework. The modified gravity theory of $f(R)$, which has gained widespread attention, presents a theoretical framework for gravitational corrections. The $f(R)$ modified gravity theory introduces a novel perspective wherein the curvature scalar $R$ is allowed to take on any arbitrary function $f(R)$ rather than being linear. By incorporating this additional degree of freedom, the $f(R)$ theory can address phenomena that are not adequately explained by general relativity \citep{CANTATA:2021ktz}, thereby providing a new theoretical framework for cosmology and cosmic evolution. 
 
 As early as the 1980s, a modified gravity model was proposed by Starobinsky to explain inflation \citep{Starobinsky:1980te}. Subsequently, with the discovery of cosmic acceleration during the late stages of the Universe \citep{SupernovaCosmologyProject:1998vns, SupernovaSearchTeam:1998fmf}, the $f(R)$ theory began to be considered as a tool for explaining this phenomenon. \cite{Basilakos:2013nfa} introduced a method for solving ordinary differential equations using series expansions to obtain the Hubble parameter, enabling a more efficient constraint of the $f(R)$ theory using cosmological observations. In terms of kinematics, \cite{Kumar:2023bqj} utilized the latest Type Ia supernova (SN Ia) data and conducted a joint analysis with baryon acoustic oscillations (BAO) and Big Bang nucleosynthesis (BBN) to provide updated observational constraints on two $f(R)$ gravity models (Hu-Sawicki and Starobinsky models). They found slight evidence for $f(R)$ gravity under the dynamics of Hu-Sawicki, but the inclusion of progenitor distances made the model compatible with general relativity. \cite{Dainotti:2023fha} performed a binning analysis of PantheonPlusSH0ES, obtaining different values of $H_0$, and proposed that $H_0$ undergoes a slow decline with $z$, speculating that the $f(R)$ modified gravity theory is an effective model for explaining this trend.
 \cite{Qi:2023ncd} studied the late-time dynamics of the Universe under the $f(R)$ model and obtained feasible late-time cosmological models through parameter tuning. They compared these models with the $\Lambda \text{CDM}$ model using SN Ia data and found good agreement between theory and data.

Undoubtedly, future SN Ia and BAO observations will provide more stringent constraint on the $f(R)$ theory, such as the Legacy Survey of Space and Time (LSST) \citep{LSSTDarkEnergyScience:2012kar}, Euclid \citep{Euclid:2023tqw}, Dark Energy Spectroscopic Instrument (DESI) \citep{casas2023euclidconstraintsfrcosmologies}, etc. The China Space Station Telescope (CSST) is a next-generation Stage~IV 2-meter sky survey telescope. It is designed for simultaneous photometric imaging and slitless grating spectroscopic measurements. Over approximately 10 years of observation, the CSST will cover a sky area of $17500\, \text{deg}^2$, with a field of view (FOV) of $1.1\, \text{deg}^2$. Its wavelength coverage ranges from near-ultraviolet to near-infrared, with seven photometric and three spectroscopic bands. Besides, the CSST also can perform 9 deg$^2$ ultra-deep field (UDF) survey for observing high-$z$ galaxies and SNe Ia \citep{Wang:2024slm}. Therefore, through the observations of weak gravitational lensing, galaxy clustering, SN Ia, and other cosmological probes, the CSST can reconstruct the history of cosmic expansion and structure growth with high precision, and hence provide accurate constraints on the modified gravity models, enabling a rigorous distinction between dark energy and modified gravity theories on cosmological scales.

In this study, we predict the constraint on different $f(R)$ theories using the mock data of the CSST SN~Ia \citep{Wang:2024slm} and BAO \citep{Miao:2023umi} observations.  All the simulations were obtained based on the flat Universe of \cite{Planck:2018vyg}, and the fiducial values of our cosmological parameters were set to $[h,\Omega_{m0},\Omega_{b0}, b, A_s, n_s, N_{\text{eff}}] = [0.673,0.313,0.0049, 0, 2.099\times 10^{-9}, 0.965, 3.04]$, where $b$ is a parameter introduced by the modification of the gravitational theory by $f(R)$. Our approach involves a comprehensive analysis that integrates these observational datasets to refine and narrow down the permissible parameter space within the context of the modified gravity theory $f(R)$.
 The structure of this paper is as follows: we introduce the basic cosmological theory related to the $f(R)$ theory and the method of obtaining the Hubble parameters under f(R) models in Section \ref{sect:Cosmo}; in Section \ref{sect:data} we discuss the relevant mock data we use; in Section \ref{sec:Bayesian} we show the parameter constraint and the model comparison methods used in this work. We give the results and summary in Section \ref{sect:result} and \ref{sec:conclusion}.

\section{Cosmology of $f(R)$ Theory}
\label{sect:Cosmo}
\subsection{$f(R)$ basics}

By modifying the Einstein-Hilbert action of General Relativity, the $f(R)$ theory can be derived by \citep{Nojiri_2011,Nojiri_2017}
\begin{equation}\label{eq1}
    S = \int d^4 x \sqrt{-g} \left( \frac{f(R)}{16 \pi G } + \mathcal{L}_{m} + \mathcal{L}_{r}\right).
\end{equation}
Here $f(R)$ denotes the function of Ricci scalar $R$, and $\mathcal{L}_{m}$ and $\mathcal{L}_{r} $ represent the Lagrangian densities for matter and radiation, respectively. The field equations for the $f(R)$ theory are obtained by variating Eq. (\ref{eq1}), and we have
\begin{equation}\label{eq2}
    f_{R} G_{a b} - \frac{1}{2} g_{a b} f + \frac{1}{2} g_{a b} f_{R} R - \nabla_{a} \nabla_{b}  f_{R} + g_{a b}  \Box  f_{R}  
    =8 \pi G [T^{(m)}_{a b} + T^{(r)}_{a b}],
\end{equation}
where $f$ is the simple form of $f(R)$, $f_{R}$ denotes the first derivative of $f(R)$ with respect to $R$. We assume that the presence of an ideal fluid in the Universe is composed of cold dark matter and radiation. $T^{(m)}_{a b} $ and $ T^{(r)}_{a b}$ represent the energy-momentum tensors for the matter sector and the radiation sector, respectively. 
For a spatially flat universe and assuming the Friedmann-Lemaître-Robertson-Walker (FLRW) metric, Eq. (\ref{eq2}) gives
\begin{align}
    3 f_{R} H^2 =& 8 \pi G (\rho_{m} + \rho_{r}) + \frac{1}{2}(f_{R} R - f) - 3H\dot{f}_{R} \label{eq6},\\
    - 2 f_{R} \dot{H} =& 8 \pi G (\rho_{m} + p_{m} + \rho_{r} + p_{r}) + \ddot{f}_{R} - H \ddot{f}_{R}.\label{eq7}
\end{align}
Here $H$ is the Hubble parameter, $\rho_x$ and $p_x$ are the energy density and pressure for matter or radiation, respectively.
We can define the effective energy density and pressure $\rho_{\text{eff}} $ and $p_{\text{eff}}$ as \citep{DeFelice:2010aj}
\begin{equation}\label{eq8}
    \rho_{\text{eff}} \equiv \frac{1}{8 \pi G} \left( \frac{1}{2} (f_{R} R - f)- 3H \dot{f}_{R} + 3(1- f_{R}) H^2 \right),
\end{equation}
\begin{equation}\label{eq9}
    p_{\text{eff}} \equiv \frac{1}{8 \pi G} \biggl[ -\frac{1}{2} (f_{R} R - f) -(1 -f_{R})(2\dot{H} + 3H^2) 
     + \ddot{f}_{R} + 2H \dot{f}_{R} \biggr].
\end{equation}
Then we obtain the modified Friedmann’s equations in the $f(R)$ theory
\begin{equation}\label{eq10}
    3f_{R}H^2 = 8\pi G ( \rho_{m} + \rho_{r} + \rho_{\text{eff}} ) + \ddot{f}_{R} - H \ddot{f}_{R},
\end{equation}
\begin{equation}\label{eq11}
    2f_{R}\dot{H} = - 8 \pi G (\rho_{m} + \frac{4}{3}\rho_{r}).
\end{equation}
The effective equation of state of the $f(R)$ gravity can be written as
\begin{equation}\label{eq12}
    w_{\text{eff}} \equiv \frac{p_{\text{eff}}}{\rho_{\text{eff}}}  = -1 - \frac{H \ddot{f}_{R} + 2 \dot{H} - 2 f_{R} \dot{H} - \ddot{f}_{R}}{\frac{1}{2}(f_{R} R -f)- 3 H \ddot{f}_{R} +3(1 - f_{R}) H^2}.
\end{equation}
We can easily find the deviation of $w_{{\text{eff}}}$ from $-1$ due to the modified gravitational theory in this form. 

The current observations of the Cosmic Microwave Background (CMB) have validated the reliability of the \text{$\Lambda$CDM} cosmological model in the high-redshift regime. Consequently, the cosmology under the $f(R)$ theory is expected to closely approximate the \text{$\Lambda$CDM} cosmology at high redshifts. Simultaneously, the \text{$\Lambda$CDM} model successfully predicts the phenomenon of late-time cosmic acceleration. Hence, 
the universe described by the $f(R)$ theory should also exhibit accelerating expansion at low redshifts without introducing a true cosmological constant. The requirements mentioned above can be summarized as follows \citep{Hu:2007nk}
\begin{align}\label{eq13}
	\lim_{R\to \infty}f(R)&=R - 2\Lambda,\nonumber \\
	\lim_{R\to 0} f(R)&=R.
\end{align}

A $f(R)$ model also needs to avoid several problems such as matter instability \citep{Faraoni:2006sy},
the instability of cosmological perturbations \citep{Bean:2006up}, the absence of the matter \citep{Chiba:2006jp}
era  and the inability to satisfy local gravity constraints \citep{Nojiri:2006ww}, thus viable $f(R)$ models must satisfy the following conditions \citep{Starobinsky:2007hu,Basilakos:2013nfa}
\begin{align}
    f_{R}>0 \ \text{and} \ f_{RR}>0,\ \text{for} \ R \ge &R_{0}(>0),\label{eq14}
    \\ 
    0<\left(\frac{R f_{RR}}{f_{R}} \right)_{r=-2}< 1,&\label{eq15}
\end{align}
where $f_{RR}$ denotes the second derivative of $f(R)$ with respect to $R$, $R_{0}$ is the value of $R$ today and $r\equiv -R f_{R}/f$. Remindly, if the final attractor is a de Sitter point, we also need $f_R>0$ for $R \ge R_{1}(>0)$, where $R_{1}$ is the Ricci scalar at the de Sitter point.

\subsection{$f(R)$ models}
The current $f(R)$ models usually can be equivalent to the perturbations of the $\Lambda \text{CDM}$ theory, so their general form can be written in the following parameterized form
\begin{equation}\label{eq16}
    f(R) = R - 2 \Lambda y(R,b),
\end{equation}
where $\lim_{R\to \infty} y(R,b) = 1$ and $\lim_{R\to 0} y(R,b) = 0$ to satisfy the conditions presented in Eq. (\ref{eq13}).

\cite{Hu:2007nk} proposed a $f(R)$ model that accelerates the cosmic expansion without a cosmological constant, and satisfies both cosmological and solar-system tests in the small-field limit of the parameter space, which takes the form as
\begin{equation}\label{eq17}
    f(R) = R - m^2\frac{c_1 (R/m^2)^n}{c_2 (R/m^2)^n + 1},
\end{equation}
where $m^2 \equiv \kappa^2 \bar{\rho}_0/3$ relates to $\kappa^2=1/16\pi G$, the average density today $\bar{\rho}_0$, and $c_1$ and $c_2$ as dimensionless parameters. In the study by \cite{Capozziello:2007eu}, it was suggested that $n$ is an integer. Therefore, for simplicity, we consider the case where $n = 1$, and then we can rewrite Eq. \ref{eq17} in a parameterized form as Eq.\ref{eq16}:
\begin{equation}\label{eq18}
    f_{\tiny{\mathrm{HS}}}(R) = R -\frac{2\Lambda}{1 + \left(\frac{b\Lambda}{R}\right)^n} ,
\end{equation}
where $\Lambda = m^2 c_1/c_2$, $b=2 c_2^{1-1/n}/c_1$ and $y_{\mathrm{HS}}(R,b) = \left[1 + \left(\frac{b\Lambda}{R}\right)^n\right]^{-1}$. It can be noted that when $b \to 0$ (or equally $c_1 \to \infty$), we have $f(R) \to R - 2\Lambda$, i.e. the Hu-Sawicki model returns to the $\Lambda \text{CDM}$ model.


Besides, \cite{Starobinsky:2007hu} also proposed an $f(R)$ model, which is given by
\begin{equation}\label{eq_St}
    f_{\mathrm{St}}(R) = R - c_1 m^2 \left[ 1 - \frac{1}{(1+R^2/m^4)^{n}} \right].
\end{equation}
Similarly, when $\Lambda = \frac{c_1 m^2}{2}$, $b = \frac{2}{c_1}$ and $y_{\mathrm{St}}(R,b) = 1- \left( 1 + \left(\frac{R}{b \Lambda}\right)^2 \right)^{-n}$, Eq. (\ref{eq_St}) can be rewritten as
\begin{equation}\label{eq_St_para}
    f_{\mathrm{St}}(R) = R - 2\Lambda \left\{1- \left[ 1 + \left(\frac{R}{b \Lambda}\right)^2 \right]^{-n} \right\}.
\end{equation}
We can find that the Starobinsky model will return to the $\Lambda\text{CDM}$ model when $b = 0$.

In addition, an alternative parameterization is also mentioned in \cite{Perez-Romero:2017njc}, which can be expressed by
\begin{equation}\label{eq_para_2}
    f(R) = R - \frac{2\Lambda}{1+ b p(R,\Lambda)}.
\end{equation}
This model under this parameterization method also yields an expansion history similar to the expansion history of $\Lambda \text{CDM}$, as well as a specific expression for the Hubble parameter via the modified Friedman equation.
Here we use a parameterized model of $f(R)$ in which the form $p(R,\Lambda)$ is $\mathrm{ArcTanh} (R,\Lambda)$, i.e.
\begin{equation}\label{eq_AcT}
    f_{\mathrm{AcT}}(R) = R - \frac{2\Lambda}{1 + b \mathrm{ArcTanh}(\Lambda/R)}.
\end{equation}
It can be seen that when $b = 0$, the ArcTanh model will also return to the $\Lambda \text{CDM}$ model. We will discuss the constraints on these three $f(R)$ models in the CSST SN Ia and BAO surveys.

Besides, in other studies \citep[e.g.][]{Liu:2016xes, Koyama_2016}, the background strength of modified gravity, $f_{R0}$, is also employed to characterize the difference between modified gravity and general relativity, where $f_{R0}$ is defined as $(\frac{df}{dR}-1)|_{z=0}$. For convenience, $f_{R0}$ is often expressed as a logarithmic function of $|f_{R0}|$, i.e. $\log_{10}|f_{R0}|$. When $\log_{10}|f_{R0}|$ is smaller, the strength of modified gravity is weaker. The relationship between $\log_{10}|f_{R0}|$ and $b$ is dependent on the specific model employed. For the three models utilized in this study, $\log_{10}|f_{R0}|$ is a function of $b$,  $\Lambda$ and $R$. The closer $b$ is to $0$, the smaller $\log_{10}|f_{R0}|$ is, when fixing $\Lambda$ and $R$.

\subsection{Hubble parameter in $f(R)$ theory}
Our investigation focuses on the cosmic late-time accelerating expansion phenomenon in the framework of the $f(R)$ theory, so it is necessary to obtain the corresponding Hubble parameters at different redshifts in this theory. It is noted that Eq.~(\ref{eq6}) represents a fourth-order ordinary differential equation (ODE) with respect to the Hubble parameter. In principle, we can obtain the solution for the Hubble parameter at a given redshift by solving the ODE. However, this method yields a highly complex solution, giving rise to various issues during the computational process, such as difficulties in integration using standard methods. To avoid these issues, \cite{Basilakos:2013nfa} introduced a perturbation method that involves expanding the Hubble parameter in $f(R)$ theory around the vicinity of the Hubble parameter in the \text{$\Lambda$CDM} model. To facilitate the use of the perturbation method described above, Eq. (\ref{eq6}) can be reformulated as follows 
\begin{equation}\label{eq23}
    -f_{R} H^2(z) + \Omega_{m0} (1+z)^3 + \Omega_{r0}(1+z)^4 + \frac{f_{R}R - f}{6} = f_{RR}H^2(z)R'(z).
\end{equation}
Subsequently, we perform a perturbative expansion of $E(z)=H(z)/H_0$ for the Hu-Sawicki model, Starobinsky model, and ArcTanh model around $b=0$, respectively. Then we have
\begin{align}
    E^2_{\text{HS}}(z) = \frac{H^2_{\text{HS}}(z)}{H^2_0} = E^2_{\Lambda}(z) + b \delta E^2_{1,\text{HS}}(z) + b^2 \delta E^2_{2,\text{HS}}(z) + \mathcal{O}(2),\label{eq24} \\
    E^2_{\text{St}}(z) = \frac{H^2_{\text{St}}(z)}{H^2_0} = E^2_{\Lambda}(z) + b^2 \delta E^2_{1,\text{St}}(z) + b^4 \delta E^2_{2,\text{St}}(z) + \mathcal{O}(2),\label{eq25} \\
    E^2_{\text{AcT}}(z) = \frac{H^2_{\text{AcT}}(z)}{H^2_0} = E^2_{\Lambda}(z) + b \delta E^2_{1,\text{AcT}}(z) + b^2 \delta E^2_{2,\text{AcT}}(z) + \mathcal{O}(2),\label{eq26} 
\end{align}
where $E_{\Lambda}(z)$ is the standardized Hubble parameters in the \text{$\Lambda$CDM} model, which is given by 
\begin{equation}\label{eq27}
    E^2_{\Lambda}(z) = \frac{H_{\Lambda}^2(z)}{H^2_0} = \Omega_{m0} (1+z)^3 + \Omega_{r0}(1+z)^4 + (1 - \Omega_{m0} - \Omega_{r0}).
\end{equation}
Balancing computational precision and efficiency, perturbations are typically considered up to the second order \cite{Basilakos:2013nfa}. Since we mainly study the history of expansion during the matter-dominated period, for simplicity, we approximate $\Omega_{r0} = 0$ in our analysis. The specific forms of Eq.~(\ref{eq24}), Eq.~(\ref{eq25}) and Eq.~(\ref{eq26})  have been obtained in \cite{Sultana:2022qzn} and shown in Appendix \ref{sec:Hz}.

\section{Mock Data}
\label{sect:data}
\subsection{Type Ia supernovae}
SN Ia, serving as cosmic standard candles, are crucial in establishing the standard cosmological model. The measurement of the distance modulus of SN Ia can effectively determine the luminosity distance $d_{L}$ at a given redshift, limit the slope of the late-time expansion rate, and consequently constrain the cosmological parameters.
The CSST-UDF survey is expected to cover a sky area of $9$ square degrees with 250 s $\times$ 60 exposures in two years, reaching a survey depth of approximately $i=26$ AB mag for 5$\sigma$ point source detection in one exposure. 

\begin{figure}
    \centering
    \includegraphics[width = \linewidth]{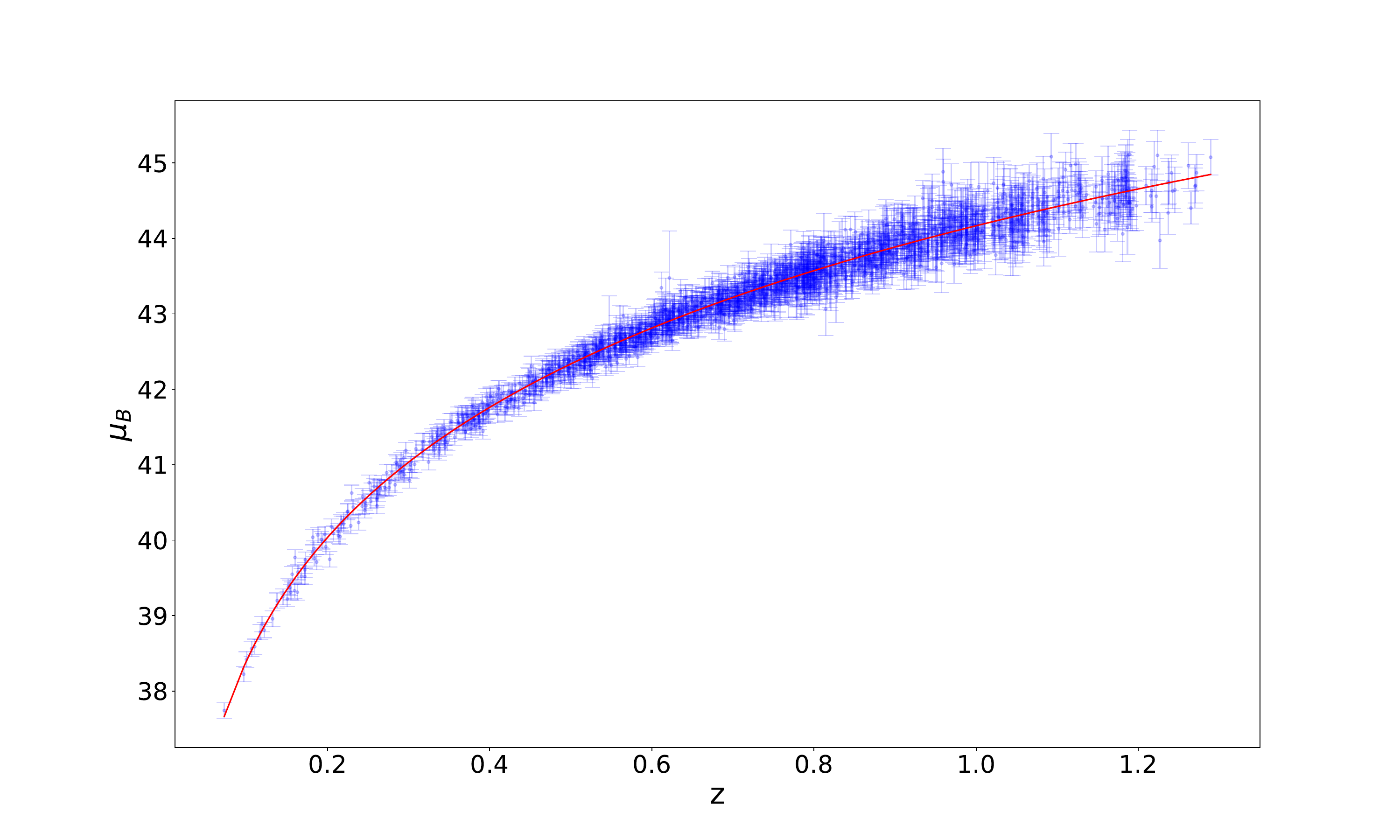}
    \caption{The mock data of distance modulus as a functin of the input redshift for the SN Ia survey in the CSST-UDF. The red solid curve represents the theoretical distance modulus using the fiducial values of the cosmological parameters and nuisance parameters in the SN Ia model.}
    \label{fig_SNIa}
\end{figure}

\cite{Wang:2024slm} utilized the SALT3 \citep{Kenworthy_2021} model and associated supernova spectral energy distribution (SED) templates to generate mock light curves of SNe Ia and different types of core-collapse supernovae (CCSNe) for the CSST-UDF survey.  Using the fitting results of mock SN Ia light curves, the SN Ia distance modulus at a given redshift can be derived by 
\begin{equation}
    \mu_B = m_B - M_B + \alpha x_1 - \beta c,
\end{equation}
where $m_B$ and $M_B$ are the $B$ band apparent and absolute magnitudes respectively, and $x_1$ and $c$ are the light-curve parameters related to time-dependent variation and color. We can obtain the photometric redshift $z$, $m_B$, $x_1$ and $c$ from the light-curve fitting process, and $M_B$, $\alpha$ and $\beta$ are the nuisance parameters and set to be free parameters when fitting the  cosmological parameters. 

Folowing \cite{Wang:2024slm}, we generate the SN Ia mock data for the CSST-UDF survey based on the cosmological parameters from Planck 2018 to constrain the $f(R)$ theory, which contains about 1897 SNe Ia in the redshift range from $z=0$ to 1.2. The fiducial values of the nuisance parameters are set to be $\alpha =0.16$, $\beta = 3.0$, and $M_B=-19.25$. In Fig.~\ref{fig_SNIa}, we show the Hubble diagram as a function of the input redshift for the SN Ia mock data. We find that, as expected, the CSST-UDF survey can obtain large fraction of high-$z$ SNe Ia, which are about 80\% and 15\% of the total SN Ia sample at $z>0.5$ and $z>1$, respectively. Note that we do not consider the contamination of CCSNe in the  fitting process of the cosmological parameters for simplicity, since it can be effectively suppressed in the data analysis and will not affect the results \citep{Wang:2024slm}.

\subsection{Baryon Acoustic Oscillation}

\begin{figure}[t]
  \begin{subfigure}[t]{0.5\linewidth}
  \centering
   \includegraphics[width=0.93\linewidth]{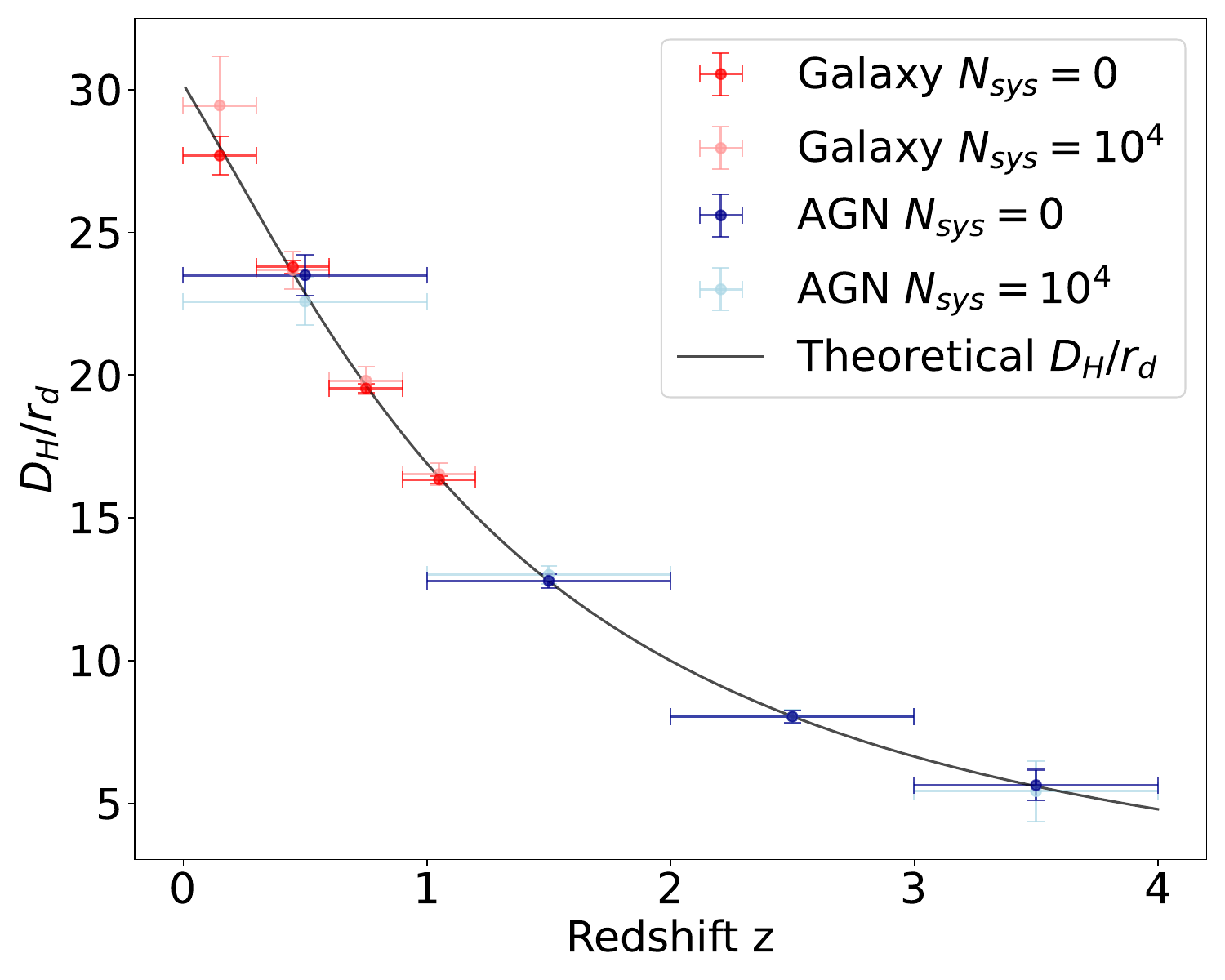}
	  \caption{\label{Fig2.1}{$D_H/r_d$} vs Redshift $z$} 
  \end{subfigure}%
  \begin{subfigure}[t]{0.5\textwidth}
  \centering
   \includegraphics[width=0.93\linewidth]{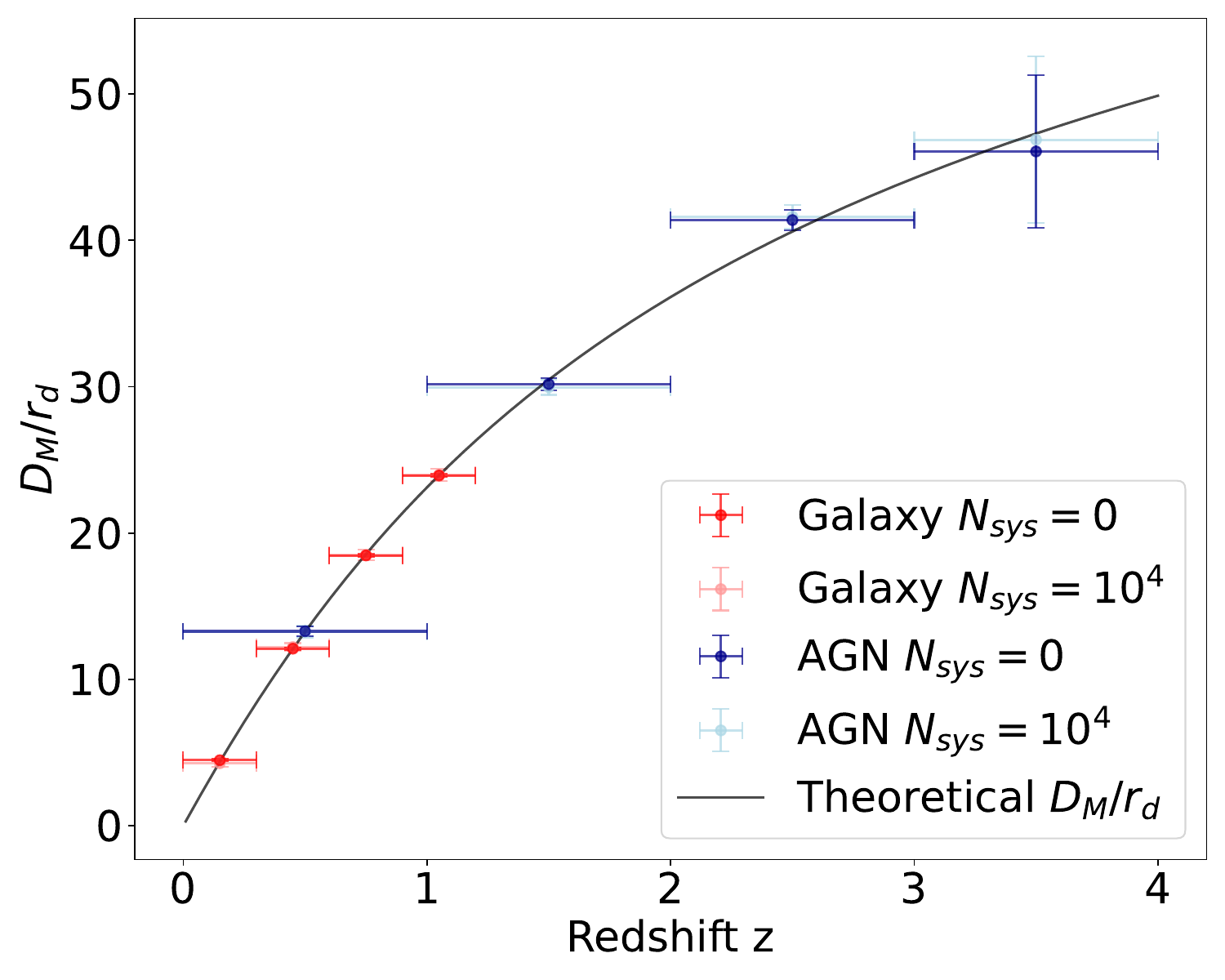}
	  \caption{\label{Fig2.2}{$D_M/r_d$ vs Redshift $z$}}
  \end{subfigure}%
  \caption{The BAO mock data of the CSST slitless spectroscopic galaxy (red) and AGN (blue) surveys from \cite{Miao:2023umi}. The data of $D_H/r_d$ and $D_M/r_d$ have been shown in the left and right panels, respectively. We also consider the systematical error $N_{\mathrm{sys}} = 0$ and $10^4\ h^{-3} {\rm Mpc^{3}}$ as the optimistic and pessimistic cases. The solid black curves denote the theoretical curves assuming the fiducial cosmological parameters.}
  \label{fig_DH_DM}
\end{figure}

We also make use of the BAO data given by \cite{Miao:2023umi} to constrain the models of the $f(R)$ theory. The BAO mock datasets are derived from the CSST galaxy and active galactic nucleus (AGN) spectroscopic surveys, which cover the redshift ranges $z\in(0,1.2)$ for galaxy survey and $z\in(0,4)$ for the AGN survey, respectively. To reduce the nonlinear effect, the reconstruction technique is applied in the BAO analysis for the CSST galaxy survey. In Fig.~\ref{fig_DH_DM}, we plot the BAO data used in our work. The data for the Hubble distance $D_H/r_d$ and comoving angular diameter distance $D_M/r_d$ have been shown in the left and right panels, respectively, where $r_d$ is the size of the sound horizon at the drag epoch. The BAO data are divided into four redshift bins for both CSST galaxy and AGN surveys, and set the systematic error of the calibration of the slitless spectroscopic survey $N_{\mathrm{sys}} = 0$ and $10^4\ h^{-3} {\rm Mpc^{3}}$ as the optimistic and pessimistic cases, respectively.
The BAO mock data are derived based on the $\Lambda \text{CDM}$ model, using the cosmological parameters derived from the Planck 2018 results as the fiducial values, which is the same as the SN Ia case.

\section{Model fitting and comparison}\label{sec:Bayesian}

We adopt Markov Chain Monte Carlo (MCMC) methods to constrain the cosmological parameters of the $f(R)$ models. The likelihood function can be estimated as $\mathcal{L} \propto \exp( -\frac{\chi^2}{2})$, where $\chi^2$ is the chi-square.  For SN Ia, it can be expressed as
\begin{equation}\label{eq_chiS_SNIa}
    \chi^2_{\rm SN} = \sum_{i} \frac{(\mu^i_{\rm obs}-\mu_{\rm th})^2}{\sigma_{{\rm SN},i}^2},
\end{equation}
where $\mu_{\rm obs}$ and $\mu_{\rm th}$ are the observational and theoretical distance moduli, and $\sigma_{\rm SN}$ is the data error. Since SN~Ia is produced by the explosion of white dwarfs accreting material greater than the Chandrasekhar limit, its absolute magnitude $M_{B}$ is related to the gravitational constant $G$. In the $f(R)$ model, $G$ is no longer a constant, but can vary as a function of time or redshift, i.e. $G_{f(R)}(z)$, which can be written as \citep{Kumar:2023bqj}
\begin{equation}\label{eq30}
    G_{f(R)}(z) = \frac{G}{f_R} \left(  \frac{1 + 4k^2 m /a^2}{1 + 3k^2m/a^2} \right),
\end{equation}
where $m = \frac{f_{RR}}{f_{R}}$ and $k = 0.1\ h\, {\rm Mpc^{-1}}$. Then the theoretical SN~Ia distance modulus will be corrected as \citep{Gazta_aga_2001, Wright_2018}
\begin{equation}\label{eq31}
    \mu_{f_{(R)}} = 5 \log_{10}(\frac{d_L}{Mpc}) + 25 + \frac{15}{4} \log_{10}(\frac{G_{f(R)}(z)}{G}).
\end{equation}
Here $d_L(z)$ is the luminosity distance for an object at redshift $z$, and it can be expressed by
\begin{equation}\label{eq28}
    d_L(z) = c(1+z)\int^z_0 \frac{1}{H(z')} d z'.
\end{equation}

The chi-square of BAO for both galaxy and AGN surveys is given by 
\begin{equation}\label{eq_chiS_BAO}
    \chi^2_{\rm BAO} = \chi^2_{\parallel} + \chi^2_{\perp} = \sum_{i} \frac{\left[(D_{H}/r_d)^i_{\rm obs} - (D_H/r_d)_{\rm th} \right]^2} {\sigma^2_{\parallel,i}} + 
    \sum_{i} \frac{\left[(D_{M}/r_d)^i_{\rm obs} - (D_M/r_d)_{\rm th}\right]^2}{\sigma^2_{\perp,i}} .
\end{equation}
Theoretically, the feature of BAO along the line of sight can be characterized by the Hubble distance $D_{H}(z) = c/H(z)$, and that perpendicular to the line of sight can be described by the comoving angular diameter distance:
\begin{equation}\label{eq32}
    D_M(z) = c\int^{z}_{0} \frac{d z^{'}}{ H(z^{'}) }.
\end{equation}
In theory, $r_{\text{d}}$ is related to the speed of sound $c_s(z)$ \citep{Brieden:2022heh}, which is given by
\begin{equation}\label{eq_rd}
    r_{\text{d}} = \int^{z_{\text{d}}}_{\infty} \frac{c_s(z)}{H(z)} d z
    = \frac{147.05}{Mpc} \left( \frac{\Omega_{m} h^2}{ 0.1432 } \right)^{-0.23} \left( \frac{N_{\mathrm{eff}}}{3.04} \right)^{-0.1} \left( \frac{\Omega_{b} h^2}{0.02235} \right)^{-0.13}.
\end{equation}
Here we fix the effective number of neutrino species $N_{\rm eff}=3.04$ and the baryon density $\Omega_b=0.02235\pm0.00037$ \citep{Sch_neberg_2019,Sch_neberg_2022} in the fitting process.

Then we can obtain the joint likelihood function, and it takes the form as
\begin{equation}\label{eq_L_Uinon}
    \mathcal{L}_{\mathrm{tot}}(D_{\rm tot}|\theta_{1},\theta_{2}) = \mathcal{L}_{\mathrm{SN}}(D_{\rm SN}|\theta_{1}) \times \mathcal{L}_{\rm BAO}(D_{\rm BAO}|\theta_{2}) \propto \exp[ -\frac{1}{2} (\chi^2_{\mathrm{SN}} + \chi^2_{\mathrm{BAO}} ) ],  
\end{equation}
where $D$ denotes the dataset, $\theta_1 = (\alpha,\beta,M_B,h,\Omega_{m0},b)$ and $\theta_2 = (h,\Omega_{m0},b)$. 
We employ {\tt emcee}\footnote{\url{https://github.com/dfm/emcee}} to perform the MCMC process \citep{emcee}, subsequently constraining the cosmological parameters of the three $f(R)$ models using the CSST SN Ia, BAO, and SN Ia+BAO mock data, respectively. We assume flat priors for the free paramters in the model, and we have $\alpha \in [0.11,0.17]$, $\beta \in [2.55,3.15]$, $M_B \in [-19.35, -19.15]$, $h \in [0.65,0.75]$, $\Omega_{m0} \in [0.0,0.6]$, and $b \in [-1.0,1.0]$.
We employ 30 walkers to randomly explore the parameter space for 100,000 steps. The first 100 steps are rejected as the burn-in process. After thinning the chains, we obtain about 30,000 chain points to illustrate the properbility distribution functions (PDFs) of the model parameters in each case.

We also utilize the Akaike Information Criterion (AIC), Bayesian Information Criterion (BIC), $\chi^2_{\mathrm{reduced}}$ and natural logarithm of the Bayesian evidence ($\ln{\mathcal{Z}}$) to compare the  $f(R)$ models to the $\Lambda\text{CDM}$ model. Here $\mathrm{AIC} = -2\ln{\mathcal{L}(\theta_{\mathrm{fit}})} + 2k$, $\mathrm{BIC} = -2\ln{\mathcal{L}(\theta_{\mathrm{fit}})} + k\ln{n}$, $\chi^2_{\mathrm{reduced}} = \chi^2_{\mathrm{min}}/n_{\mathrm{dof}}$ and ${\mathcal{Z}} = \int {\mathcal{L}}(D|\theta, M) P(\theta|M) d \theta$, where $\theta_{\mathrm{fit}}$ is the best-fitting value, $k$ is the number of parameters, $n$ is the number of data, $n_{\mathrm{dof}}= n-k$ is the degrees of freedom, $\chi^2_{\rm min}$ is the minimum chi-square, and $P(\theta|M)$ is the prior distribution of the paramter $\theta$ given a model $M$.

\section{Result and Discuss}\label{sect:result}

\begin{figure}[t]
  \begin{subfigure}[t]{0.33\linewidth}
  \centering
   \includegraphics[width=\linewidth]{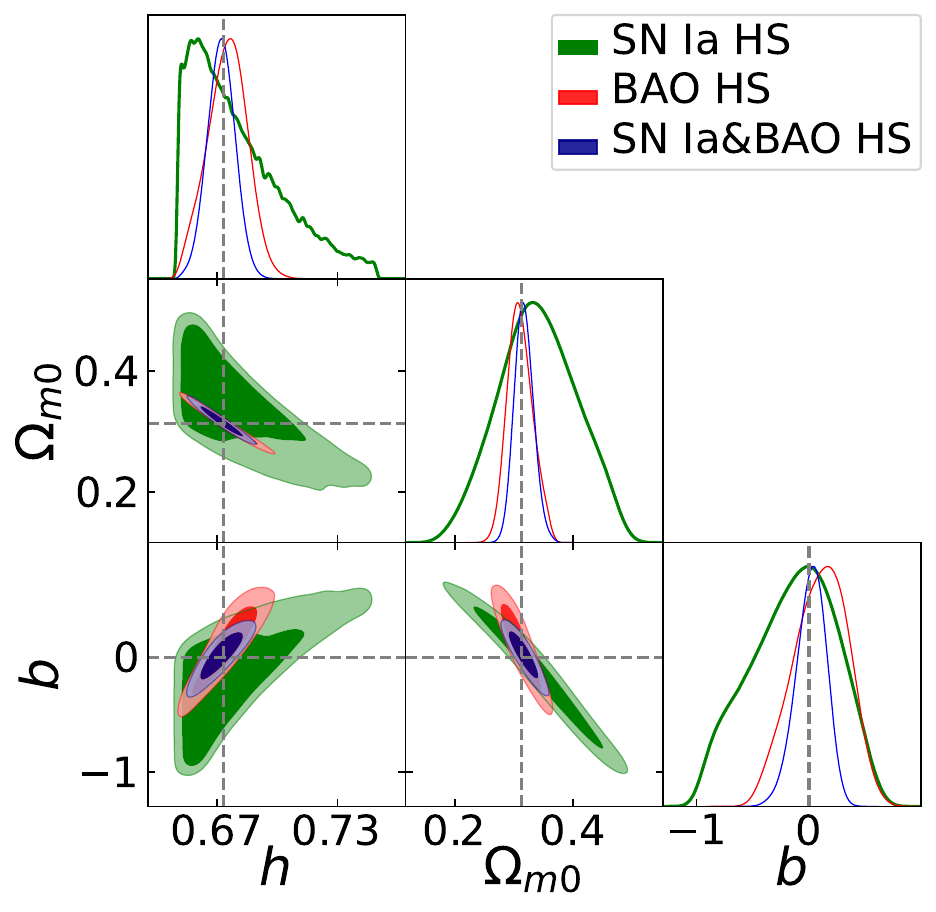}
	  \caption{\label{Fig2.1}{\small HS $N_{\mathrm{sys}} = 0$}} 
  \end{subfigure}%
  \begin{subfigure}[t]{0.33\textwidth}
  \centering
   \includegraphics[width=\linewidth]{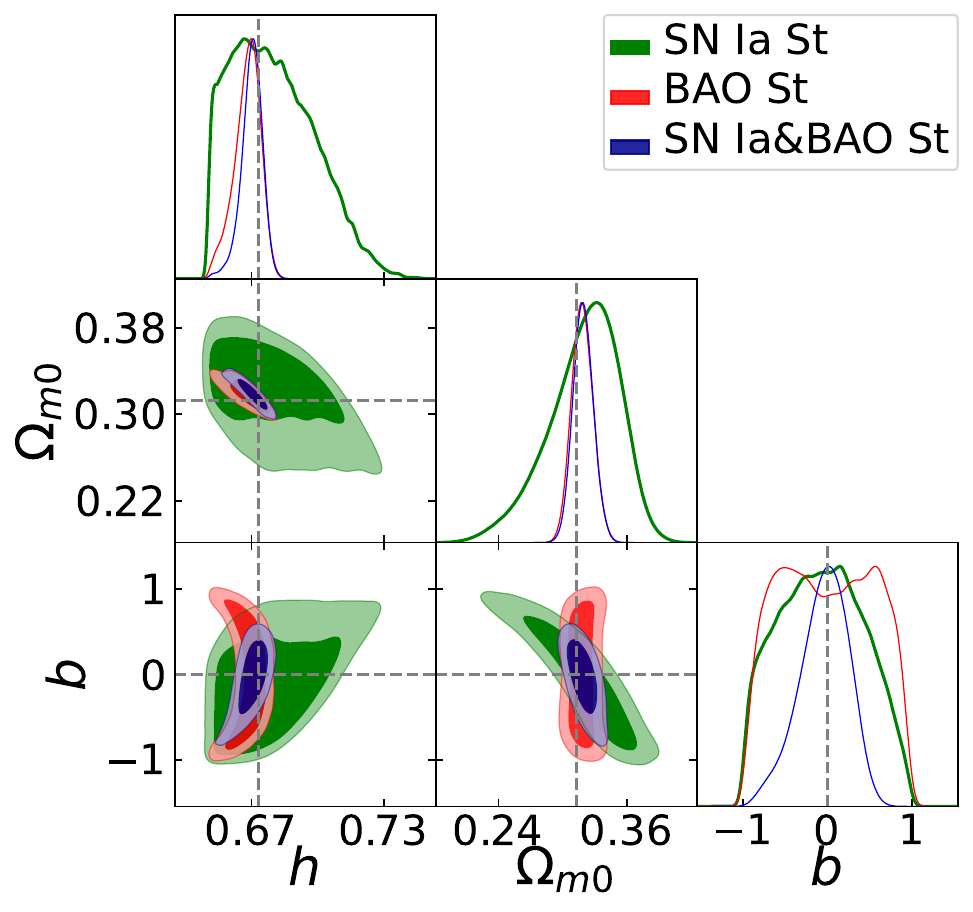}
	  \caption{\label{Fig2.2}{\small St $N_{\mathrm{sys}} = 0$}}
  \end{subfigure}%
  \begin{subfigure}[t]{0.33\textwidth}
  \centering
   \includegraphics[width=\linewidth]{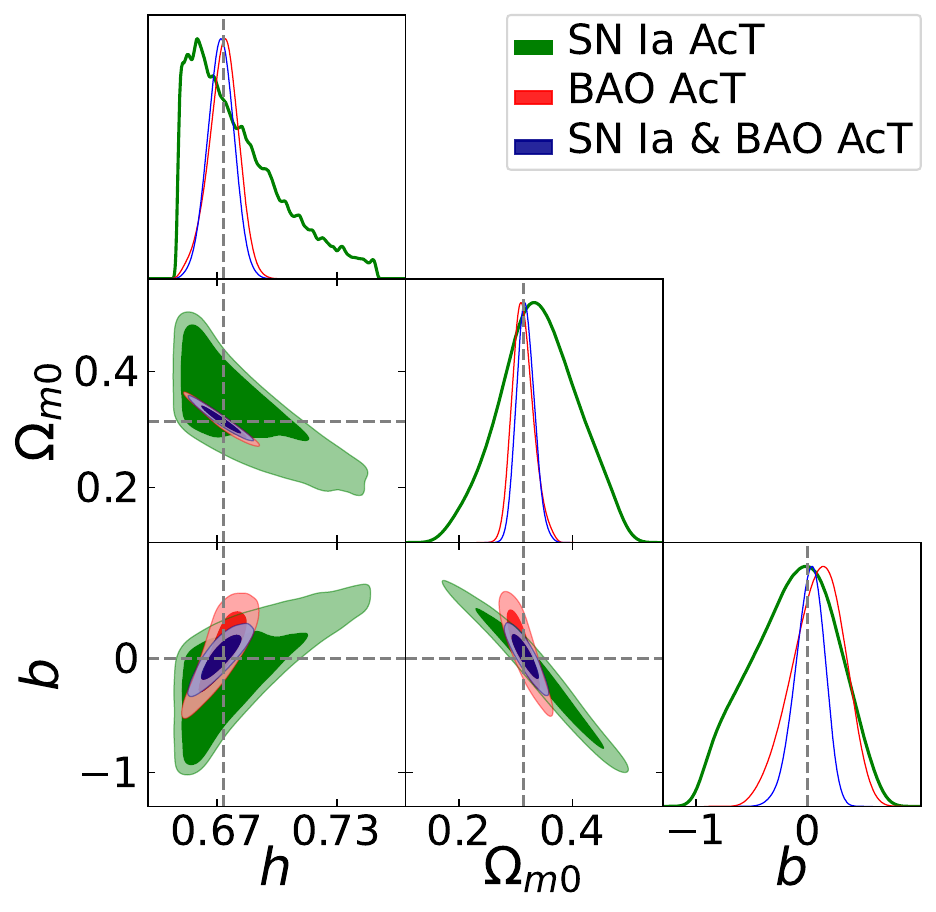}
	  \caption{\label{Fig2.3}{\small AcT $N_{\mathrm{sys}} = 0$}}
  \end{subfigure}%
  \\
   \begin{subfigure}[t]{0.33\linewidth}
  \centering
   \includegraphics[width=\linewidth]{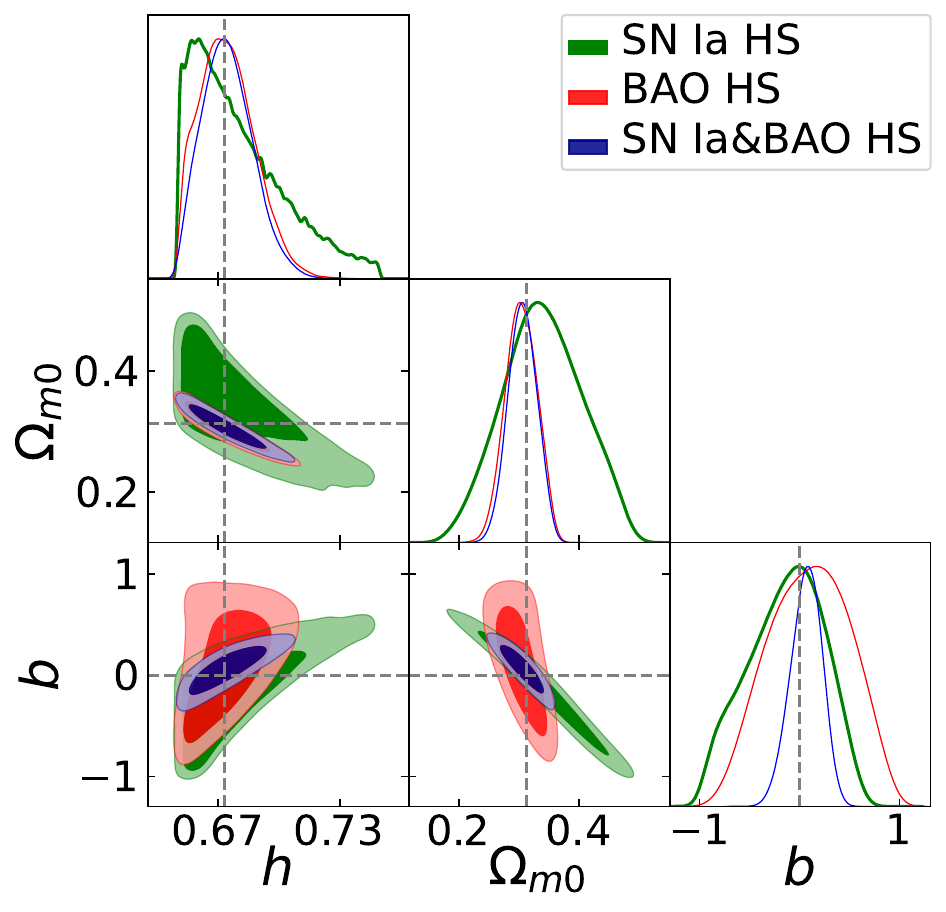}
	  \caption{\label{Fig2.4}{\small HS $N_{\mathrm{sys}} = 10^4 h^{-3}Mpc^3$} }
  \end{subfigure}%
  \begin{subfigure}[t]{0.33\textwidth}
  \centering
   \includegraphics[width=\linewidth]{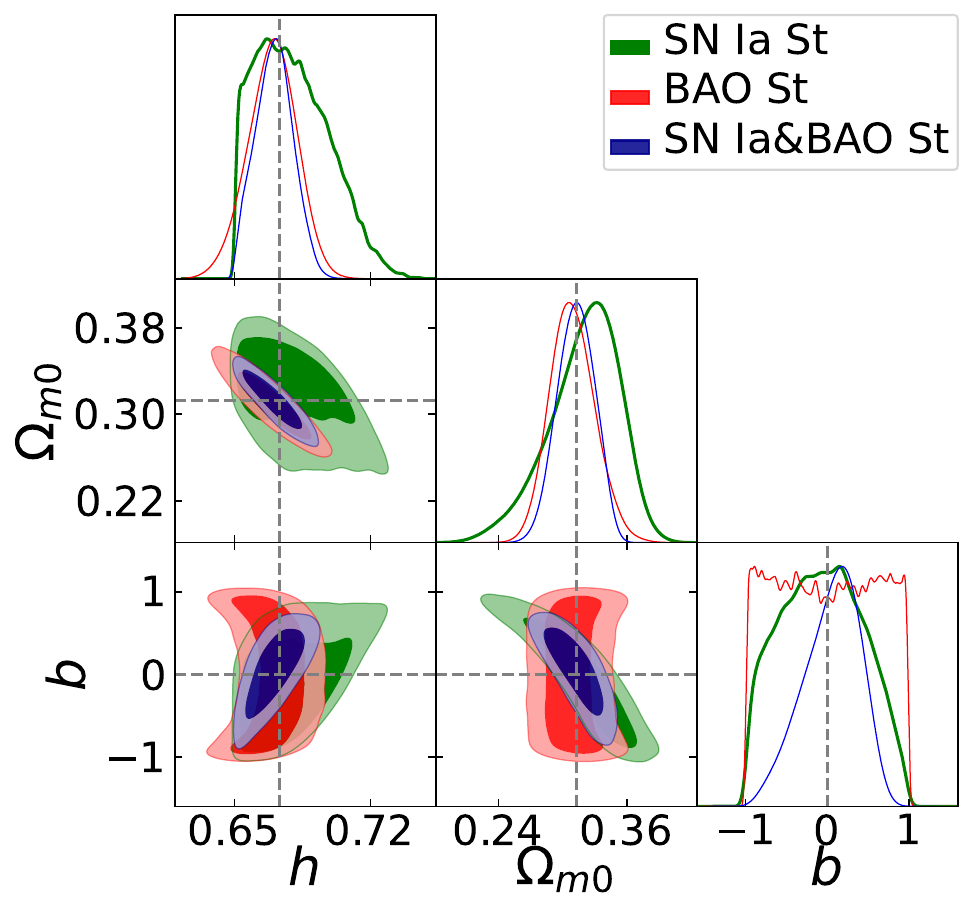}
	  \caption{\label{Fig2.5}{\small St $N_{\mathrm{sys}} = 10^4 h^{-3}Mpc^3$}}
  \end{subfigure}%
  \begin{subfigure}[t]{0.33\textwidth}
  \centering
   \includegraphics[width=\linewidth]{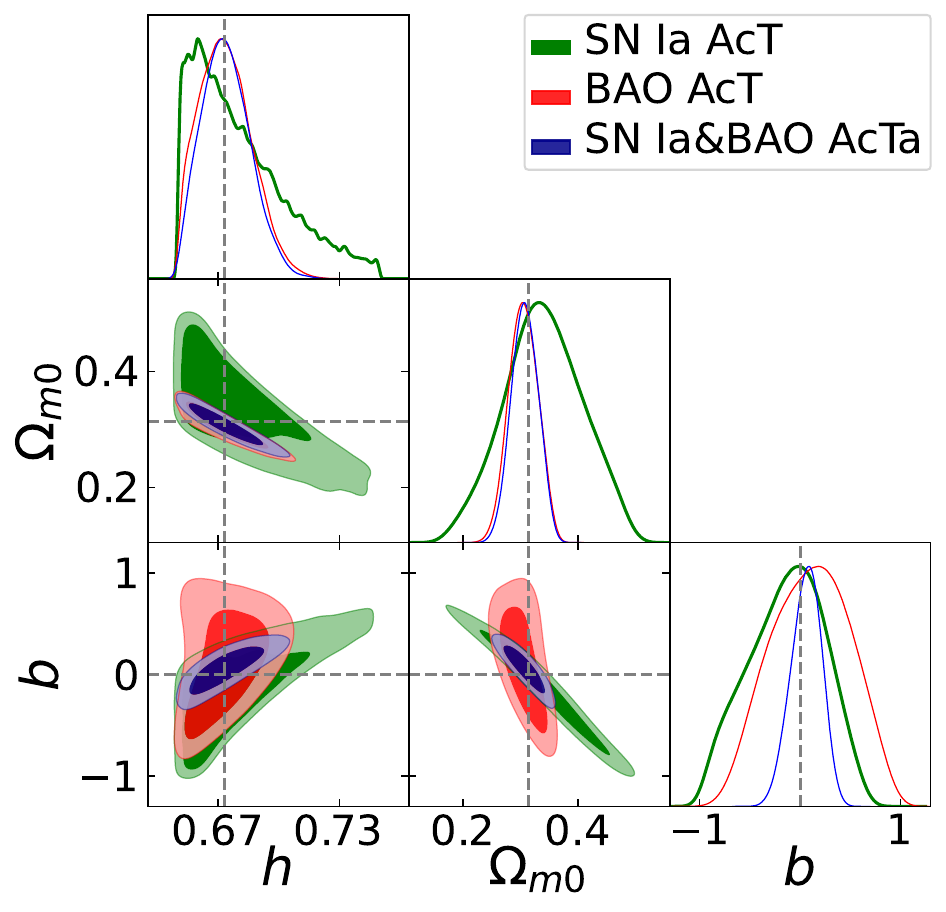}
	  \caption{\label{Fig2.6}{\small AcT $N_{\mathrm{sys}} = 10^4 h^{-3}Mpc^3$}}
  \end{subfigure}%
  \caption{The 1$\sigma$ and 2$\sigma$ contour maps and 1D PDFs of the parameters in the three $f(R)$ models for the CSST SN Ia and BAO mock data. We study the constraint results by assuming $N_{\mathrm{sys}} = 0$ (upper panels) and $10^4\ h^{-3} {\rm Mpc^{3}}$ (lower panels) for the BAO data. The dashed lines represent the fiducial parameter values.}
  \label{fig_triangleplot}
\end{figure}

\begin{table}[h]
    \centering
    \caption{The best-fits and 1$\sigma$ errors of the parameters in the $\Lambda\text{CDM}$ and three $f(R)$ models constrained by the CSST SN Ia and BAO mock data.}
    \begin{tabular}{ccccc}
       \toprule
       Dateset & Model & $h$ & $\Omega_{m0}$ & $b$  \\
       \midrule
       \multirow{4}{*}{$\text{SN Ia}$} 
       &$\Lambda\text{CDM}$&$0.685^{+0.014}_{-0.025}(2.88\%)$&$0.317^{+0.016}_{-0.010}(4.15\%)$&-\\
       &$\text{Hu-Sawicki}$& $0.661^{+0.044}_{-0.003}(3.52\%)$ & $0.330^{+0.081}_{-0.057}  (20.83\%)$ & $ 0.086^{+0.169}_{-0.643} $ \\
       &$\text{Starobinsky}$& $0.666^{+0.034}_{-0.006}(3.00\%)$ & $0.330^{+0.022}_{-0.044} (9.96\%)$ & $0.010^{+0.466}_{-0.628}$\\
       &$\text{ArcTanh}$& $0.659^{+0.045}_{-0.002}(3.55\%)$ & $0.334^{+0.078}_{-0.063}  (21.00\%)$ & $ 0.026^{+0.232}_{-0.578}$  \\
       \midrule
       \multirow{4}{*}{$\text{BAO}(N_{\text{sys}} = 0)$} 
       &$\Lambda\text{CDM}$& $0.672\pm0.004 (0.62\%)$ & $0.317^{+0.010}_{-0.008}(2.86\%)$ &-\\
       &$\text{Hu-Sawicki}$ & $0.677^{+0.008}_{-0.012}(1.49\%)$ & $0.304^{+0.029}_{-0.015} (7.09\%)$ & $0.153^{+0.195}_{-0.301}$\\
       &$\text{Starobinsky}$& $0.671^{+0.003}_{-0.009}(0.88\%)$ & $0.318\pm 0.010(3.10\%)$ & $0.001^{+0.618}_{-0.616}$\\
       &$\text{ArcTanh}$& $0.675^{+0.005}_{-0.009}(1.10\%)$ & $0.308^{+0.024}_{-0.014}(6.07\%)$ & $0.211^{+0.112}_{-0.359}$   \\
       \midrule
       \multirow{4}{*}{$\text{BAO}(N_{\text{sys}} = 10^4 
       )$} 
       &$\Lambda\text{CDM}$&$0.674^{+0.012}_{-0.009}(1.56\%)$&$ 0.298^{+0.027}_{-0.011}(6.32\%)$&-\\
       &$\text{Hu-Sawicki}$& $0.669^{+0.019}_{-0.009}(2.10\%)$ & $0.300^{+0.031}_{-0.023}(8.92\%)$ & $0.253^{+0.279}_{-0.585}$ \\
       &$\text{Starobinsky}$ & $0.670^{+0.012}_{-0.013}(1.87\%)$ & $0.302^{+0.028}_{-0.014}(6.90\%)$ & $-0.005\pm0.688$ \\
       &$\text{ArcTanh}$& $0.675^{+0.012}_{-0.015}(1.99\%)$ & $0.304^{+0.027}_{-0.024}(8.45\%)$ & $0.111^{+0.418}_{-0.424}$   \\
       \midrule
       \multirow{4}{*}{$\text{SN Ia} + \text{BAO}(N_{\text{sys}} = 0
       )$} 
       &$\Lambda\text{CDM}$&$0.671^{+0.004}_{-0.003}(0.54\%)$&$0.317^{+0.010}_{-0.006}(2.39\%)$&-\\
       &$\text{Hu-Sawicki}$ & $0.672 \pm 0.007(1.07\%)$ & $0.316^{+0.017}_{-0.015}(5.12\%)$ & $0.037^{+0.116}_{-0.153} $ \\
       &$\text{Starobinsky}$& $0.671^{+0.004}_{-0.005}(0.68\%)$ & $0.318^{+0.010}_{-0.008}(2.93\%)$ & $0.024^{+0.229}_{-0.360}$\\
       &$\text{ArcTanh}$ & $0.672^{+0.006}_{-0.007}(0.98\%)$ & $0.315^{+0.018}_{-0.014}(5.04\%)$ & $0.027^{+0.121}_{-0.141}$  \\
       \midrule
       \multirow{4}{*}{$\text{SN Ia} + \text{BAO}(N_{\text{sys}} = 10^4 
       )$} 
       &$\Lambda\text{CDM}$&$ 0.670^{+0.006}_{-0.007}(1.03\%)$&$0.312^{+0.015}_{-0.007}(3.51\%)$&-\\
       &$\text{Hu-Sawicki}$& $0.672^{+0.015}_{-0.010}(1.89\%)$ & $0.307^{+0.024}_{-0.024}(7.90\%)$ & $ 0.089^{+0.133}_{-0.181}$  \\
       &$\text{Starobinsky}$ & $ 0.671^{+0.009}_{-0.011} (1.48\%)$ & $0.312^{+0.018}_{-0.018}(5.75\%)$ & $ 0.152^{+0.246}_{-0.468}$ \\
       &$\text{ArcTanh}$ & $0.675^{+0.011}_{-0.014}(1.81\%)$ & $0.303^{+0.028}_{-0.019}(7.84\%)$ & $0.092^{+0.121}_{-0.187}$  \\
       \midrule
    \end{tabular}
    \label{Table_Cosmo_result}
\end{table}

In Fig.~\ref{fig_triangleplot}, we show the predicted 2D contour maps and 1D PDFs of the parameters in the three $f(R)$ models for the CSST SN Ia and BAO surveys. The details of the constraint results are listed in Table~\ref{Table_Cosmo_result}. Based on our utilization of observational data derived from the cosmological simulations with $b=0$, we anticipate that the constraint on $b$ should be around $0$, and similarly for $(h,\Omega_{m0})$ centering around $(0.673, 0.313)$. As we can see, Figure \ref{fig_triangleplot} shows that all parameter constraint results are closely around their fiducial values, which matches well with the expectation.

As can be seen in Figure \ref{fig_triangleplot} and Table~\ref{Table_Cosmo_result}, the constraints on $\Omega_{m0}$ and $h$ in the $\Lambda$CDM model are more stringent than that in the $f(R)$ models, since there is an additional parameter $b$ in the $f(R)$ models. The constraint results from the CSST BAO survey are basically better than the CSST-UDF SN Ia survey for these two parameters, even considering $N_{\mathrm{sys}} =10^4\ h^{-3} {\rm Mpc^{3}}$ in the BAO data. Note that the current constraint accuracy on $h$ can reach $\sim 3\%$ in the CSST-UDF SN Ia survey, and this is because we have assumed a relative narrow prior range for $M_B$, which has strong degeneracy with $h$. We can find that, the joint constraints on $\Omega_{m0}$ and $h$ in the $f(R)$ models can achieve $1\%-2\%$ and $3\%-8\%$ accuracy for the CSST SN Ia+BAO mock data.

For the $f(R)$ model parameter $b$, the constraint results are similar for the Hu-Sawicki model and the ArcTanh model, and can restrict $b$ within about $\pm0.4$ and $\pm 0.3$ for the CSST SN Ia and BAO surveys, respectively. The joint constraint can improve the result to be within $\pm0.2$ using the CSST SN Ia+BAO mock data. However, the constraints on $b$ become much worse for the Starobinsky model, which give the results within about $\pm0.6$, $\pm0.7$, and $\pm0.5$ for the SN Ia, BAO, and SN Ia+BAO data, respectively. This implies that the parameter $b$ in the Starobinsky model is not as sensitive as the other models to the SN Ia and BAO data, which is also indicated by other studies, e.g. \citep{Kumar:2023bqj,Sultana:2022qzn}.
In addition, we also explore the constraint accuracy of $\log_{10}|f_{R0}|$ corresponding to the three models based on the MCMC chains. We find that the accuracies can reach $21\%$, $33\%$, and $19\%$ for the Hu-Sawicki, Starobinsky and ArcTanh models, respectively.
Comparing our constraints to the results using the current observational data, e.g. \cite{Kumar:2023bqj}, we find that, the precision of parameter constraints on the $f(R)$ models by the CSST SN Ia+BAO dataset are comparable to or even higher than that of the eBOSS-BAO \citep{Alam_2021} + BBN \citep{Aver_2015} + PantheonPlus\&SH0ES \citep{Brout:2022vxf} dataset.

\begin{table}[h]
    \centering
    \caption{Comparison results of the Hu-Sawicki, Starobinsky, and ArcTanh models with the $\Lambda \text{CDM}$ model using the four model comparison methods of AIC, BIC, $\chi^2_{\text{reduced}}$ and $\ln{\mathcal{Z}}$ for each dataset.}
    \begin{tabular}{cccccc}
       \toprule
       Dataset & Model & $\Delta \text{AIC}$ & $\Delta \text{BIC}$ & $\Delta \chi^2_{\text{reduced}}$ & $\Delta \ln{\mathcal{Z}}$ \\
       \midrule
       \multirow{3}{*}{SN Ia} 
       &HS & 1.6577 & 7.2057 & 0.0002 & -0.0551 \\
       &St & 1.7768 & 7.3248 & 0.0003 & -0.0805 \\
       &AcT& 1.6447 & 7.1927 & 0.0002 & -0.0749 \\
       \midrule
       \multirow{3}{*}{BAO $N_{\mathrm{sys}} = 0$} 
       &HS & 1.7623 & 1.8418 & 0.1917 & -0.1539 \\
       &St & 1.8623 & 1.9417 & 0.2117 & -0.0279 \\
       &AcT& 1.7706 & 1.8500 & 0.1933 & -0.1761 \\
       \midrule
       \multirow{3}{*}{BAO $N_{\mathrm{sys}} = 10^4$} 
       &HS & 1.9841 & 2.0635 & 0.1335 & -0.1527 \\
       &St & 1.9477 & 2.0271 & 0.1262 &  0.0241 \\
       &AcT& 1.9844 & 2.0638 & 0.1336 & -0.1493 \\
       \midrule
       \multirow{3}{*}{SN Ia + BAO $N_{\mathrm{sys}} = 0$} 
       &HS & 1.9422 & 7.4944 & 0.0004 & -0.2139 \\
       &St & 2.1088 & 7.6610 & 0.0004 & -0.2297 \\
       &AcT& 1.9599 & 7.5122 & 0.0004 & -0.2165 \\
       \midrule
       \multirow{3}{*}{SN Ia + BAO $N_{\mathrm{sys}} = 10^4$} 
       &HS & 1.8254 & 7.3776 & 0.0003 & -0.1426 \\
       &St & 1.7610 & 7.3132 & 0.0003 & -0.1308 \\
       &AcT& 1.8062 & 7.3585 & 0.0003 & -0.1468 \\
       \bottomrule
    \end{tabular}
    \label{Table_Delta_AIC_BIC_lnZ}
\end{table}

We also perform model comparison with the $\Lambda$CDM model by calculating $\Delta{\mathrm{AIC}}$, $\Delta {\mathrm{BIC}}$, $\Delta {\chi^2_{\mathrm{reduced}}}$, and $\Delta \ln{\mathcal{Z}}$, and the results have been shown in Table~\ref{Table_Delta_AIC_BIC_lnZ}. As expected, all of the criteria of model comparison distinctly prefer the $\Lambda$CDM model to the $f(R)$ model, since we have assumed it as our fiducial model in the mock data generation and analysis. This indicates that the CSST SN Ia and BAO data are accurate enough to put strong constraint on the $f(R)$ theory and can distinguish it from the $\Lambda$CDM model in a high significance level. We also show the constraint results of  the nuisance parameters in the CSST-UDF SN Ia survey, i.e.  $\alpha$, $\beta$ and $M_B$ in Appendix~\ref{sec:nuisancepara}.

\section{Summary and Conclusion}\label{sec:conclusion}

In this work, we employ the simulated CSST SN Ia and BAO data to study the constraint power on the relevant parameters of the three $f(R)$ theoretical models, i.e. the Hu-Sawicki, Starobinsky and ArcTanh models.
The high-precision simulated observational data of the CSST can provide us with a good validation channel for the future constraint ability of the CSST on the $f(R)$ modified gravity theory. Firstly, following the steps outlined by \cite{Basilakos:2013nfa} and \cite{Sultana:2022qzn}, we obtained the expansion rate $H(z)$ for the three models. Then we use the CSST SN~Ia and BAO mock provided by \cite{Wang:2024slm} and \cite{Miao:2023umi} to constrain the three $f(R)$ theories. 
We find that the CSST SN Ia and BAO surveys can provide stringent constraint on the $f(R)$ models. Compared to the current results using simliar kinds of observational data, the constraints on the $f(R)$ models by the CSST SN Ia+BAO joint dataset are comparable or even higher. Besides, if considering other CSST surveys, e.g. weak gravitational lensing and galaxy clustering surveys, the constraint result can be further significantly improved. 
Therefore, we can expect that, by performing a joint analysis of these CSST cosmological probes, CSST is able to constrain and distinguish the $f(R)$ theory and the $\Lambda \text{CDM}$ theory in a high precision.

\normalem
\begin{acknowledgements}\label{sect:ac}:
J.H.Y. and Y.G. acknowledge the support from National Key R\&D Program of China grant Nos. 2022YFF0503404, 2020SKA0110402, and the CAS Project for Young Scientists in Basic Research (No. YSBR092). XC acknowledges the support of the National Natural Science Foundation of China through Grant Nos. 11473044 and 11973047, and the Chinese Academy of Science grants ZDKYYQ20200008, QYZDJ-SSW-SLH017, XDB 23040100, and XDA15020200. This work is also supported by science research grants from the China Manned Space Project with Grant Nos. CMS-CSST-2021-B01 and CMS-CSST-2021-A01.
\end{acknowledgements}

\appendix  
\section{E(z) in the three f(R) models}\label{sec:Hz}

The expressions of $E(z)$ in the Hu-Sawicki, Starobinsky and ArcTanh $f(R)$ models we use have been shown as below for reference.

Hu-Sawicki model:
\begin{equation}\label{eqA1}
\begin{aligned}
E^2_{\text{HS}}(z) =&  \frac{H_{\mathrm{HS}}(z)^2}{H_0^2}=1-\Omega_{m0}+(1+z)^3 \Omega_{m0} \\
& +\frac{6 b\left(-1+\Omega_{m0}\right)^2\left(4\left(-1+\Omega_{m0}\right)^2+(1+z)^3\left(-1+\Omega_{m0}\right) \Omega_{m0}-2(1+z)^6 \Omega_{m0}^2\right)}{(1+z)^9\left(\frac{4\left(-1+\Omega_{m0}\right)}{(1+z)^3}-\Omega_{m0}\right)^3} \\
& +\frac{b^2\left(-1+\Omega_{m0}\right)^3}{(1+z)^{24}\left(-\frac{4\left(-1+\Omega_{m0}\right)}{(1+z)^3}+\Omega_{m0}\right)^8}\left(1024(-1+\Omega_{m0})^6 \right.\\
& \left.+9216(1+z)^3\left(-1+\Omega_{m0}\right)^5 \Omega_{m0} -22848(1+z)^6\left(-1+\Omega_{m0}\right)^4 \Omega_{m0}^2 \right. \\
& \left. + 25408(1+z)^9\left(-1+\Omega_{m0}\right)^3 \Omega_{m0}^3  -7452(1+z)^{12}\left(-1+\Omega_{m0}\right)^2 \Omega_{m0}^4\right.\\
& \left.-4656(1+z)^{15}\left(-1+\Omega_{m0}\right) \Omega_{m0}^5 +37(1+z)^{18} \Omega_{m0}^6\right)
\end{aligned}
\end{equation}

\noindent Strobinsky model:
\begin{equation}\label{eqA2}
\begin{aligned}
E^2_{\text{St}}(z) = &\frac{H_{\mathrm{St}}(z)^2}{H_0^2}=1-\Omega_{m0}+(1+z)^3 \Omega_{m0} \\
& +\frac{b^2\left(-1+\Omega_{m0}\right)^3\left(32\left(-1+\Omega_{m0}\right)^2+32(1+z)^3\left(-1+\Omega_{m0}\right) \Omega_{m0}-37(1+z)^6 \Omega_{m0}^2\right)}{(1+z)^{12}\left(-\frac{4\left(-1+\Omega_{m0}\right)}{(1+z)^3}+\Omega_{m0}\right)^4} \\
& +\frac{b^4\left(-1+\Omega_{m0}\right)^5}{(1+z)^{30}\left(-\frac{4\left(-1+\Omega_{m0}\right)}{(1+z)^3}+\Omega_{m0}\right)^{10}}\left(20480\left(-1+\Omega_{m0}\right)^6\right. \\
& \left.+63488(1+z)^3\left(-1+\Omega_{m0}\right)^5 \Omega_{m0}-234880(1+z)^6\left(-1+\Omega_{m0}\right)^4 \Omega_{m0}^2\right. \\
& \left.+289024(1+z)^9\left(-1+\Omega_{m0}\right)^3 \Omega_{m0}^3 -44552(1+z)^{12}\left(-1+\Omega_{m0}\right)^2 \Omega_{m0}^4 \right.\\
& \left.- 82748(1+z)^{15}\left(-1+\Omega_{m0}\right) \Omega_{m0}^5+123(1+z)^{18} \Omega_{m0}^6\right) .
\end{aligned}
\end{equation}

\noindent ArcTanh model:
\begin{equation}\label{eqA3}
\begin{aligned}
E^2_{\text{AcT}}(z) =& \frac{H_{\mathrm{AcT}}(z)^2}{H_0^2}= 1-\Omega_{m0}+(1+z)^3 \Omega_{m0} \\
& +\frac{2 b\left(-1+\Omega_{m0}\right)^2}{3(1+z)^{15}\left(\frac{4\left(-1+\Omega_{m0}\right)}{(1+z)^3}-\Omega_{m0}\right)^5}\left(596\left(-1+\Omega_{m0}\right)^4 \right. \\
& \left.-109(1+z)^3\left(-1+\Omega_{m0}\right)^3 \Omega_{m0}-361(1+z)^6\left(-1+\Omega_{m0}\right)^2 \Omega_{m0}^2 \right. \\
& \left.+153(1+z)^9\left(-1+\Omega_{m0}\right) \Omega_{m0}^3-18(1+z)^{12} \Omega_{m0}^4 \right) \\
& +\frac{b^2 \left( -1 + \Omega_{m0} \right)^3}{9(1+z)^{36} \left( -\frac{4(-1+\Omega_{m0})}{(1+z)^3} + \Omega_{m0}\right)^{12}} \left(4189184\left(-1+\Omega_{m0}\right)^{10} \right. \\
& \left.+23851008(1+z)^3\left(-1+\Omega_{m0}\right)^9 \Omega_{m0}-95032704(1+z)^6\left(-1+\Omega_{m0}\right)^8 \Omega_{m0}^2 \right.\\
& \left.+149808704(1+z)^9\left(-1+\Omega_{m0}\right)^7 \Omega_{m0}^3-110911572(1+z)^{12}\left(-1+\Omega_{m0}\right)^6 \Omega_{m0}^4 \right. \\
& \left.+28668900(1+z)^{15}\left(-1+\Omega_{m0}\right)^5 \Omega_{m0}^5+2987537(1+z)^{18}\left(-1+\Omega_{m0}\right)^4 \Omega_{m0}^6\right. \\
& \left.-3144240(1+z)^{21}\left(-1+\Omega_{m0}\right)^3 \Omega_{m0}^7+636102(1+z)^{24}\left(-1+\Omega_{m0}\right)^2 \Omega_{m0}^8\right. \\
& \left.-47232(1+z)^{27}\left(-1+\Omega_{m0}\right) \Omega_{m0}^9+333(1+z)^{30} \Omega_{m0}^{10} \right)
\end{aligned}
\end{equation}

\section{Constraints on the nuisance parameters in the SN Ia model}\label{sec:nuisancepara}
\begin{figure}[h]
  \begin{subfigure}[t]{0.33\linewidth}
  \centering
   \includegraphics[width=\linewidth]{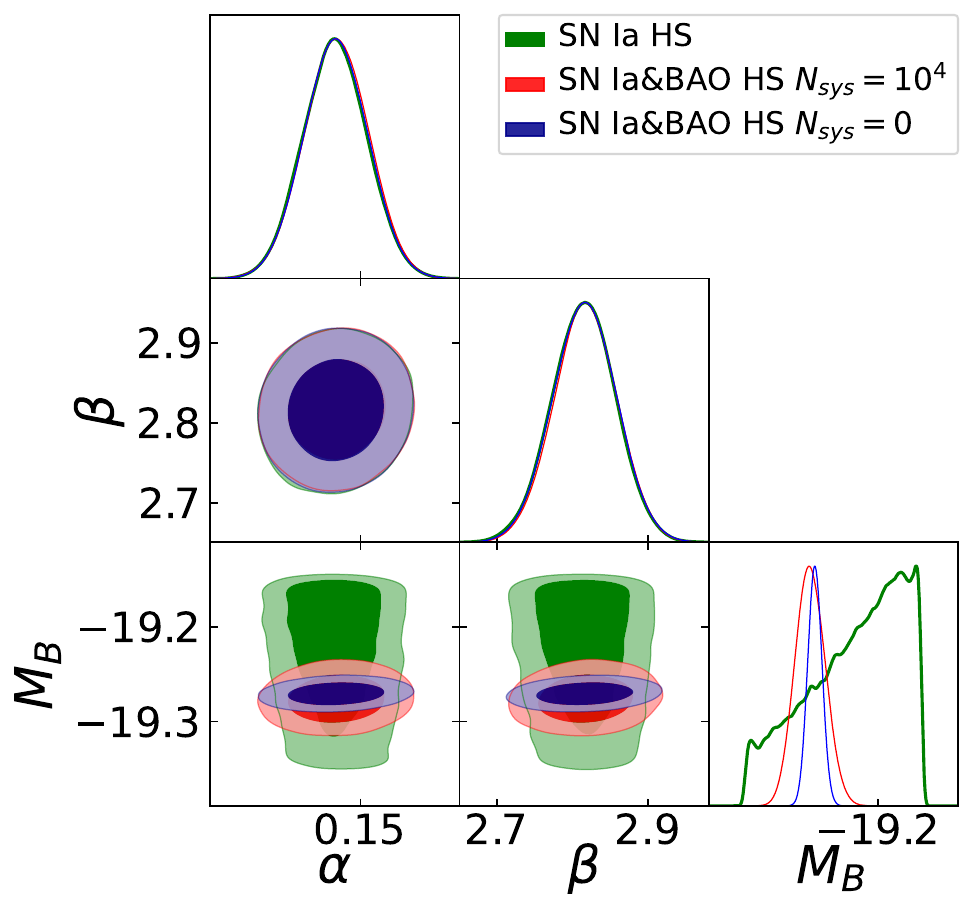}
	  \caption{\label{FigA.1}{\small Hu-Sawicki }} 
  \end{subfigure}%
  \begin{subfigure}[t]{0.33\textwidth}
  \centering
   \includegraphics[width=\linewidth]{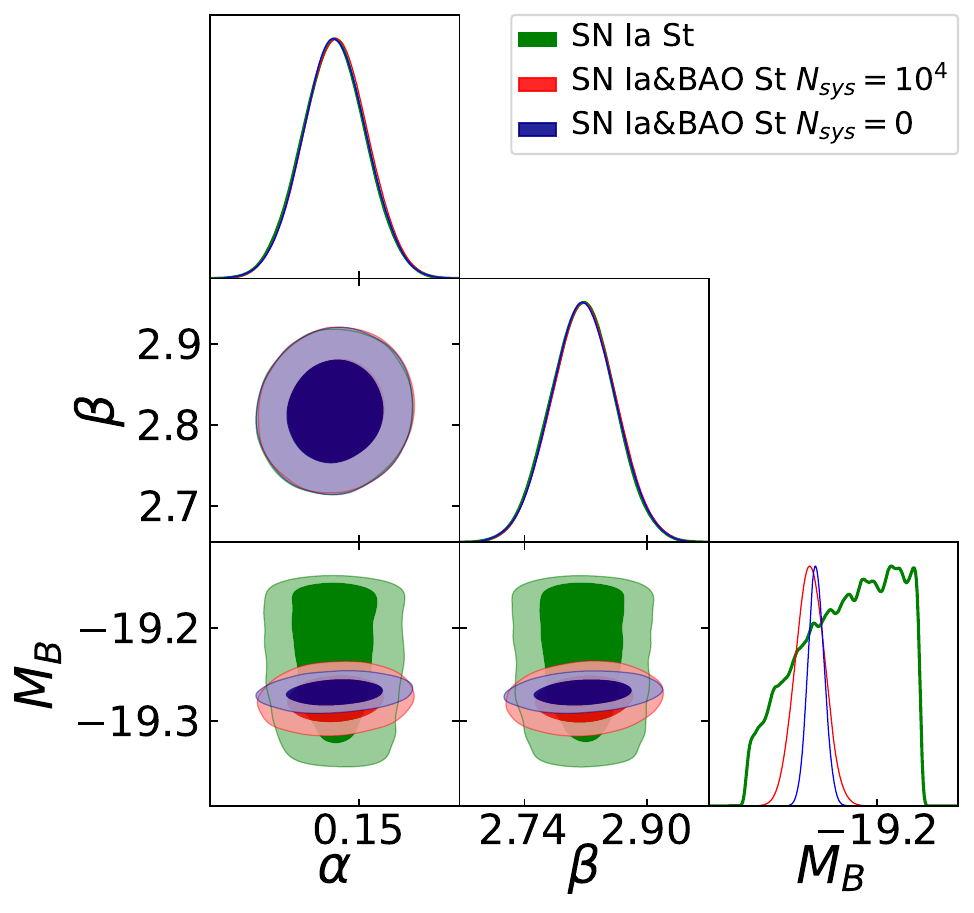}
	  \caption{\label{FigA.2}{\small Starobinsky }}
  \end{subfigure}%
  \begin{subfigure}[t]{0.33\textwidth}
  \centering
   \includegraphics[width=\linewidth]{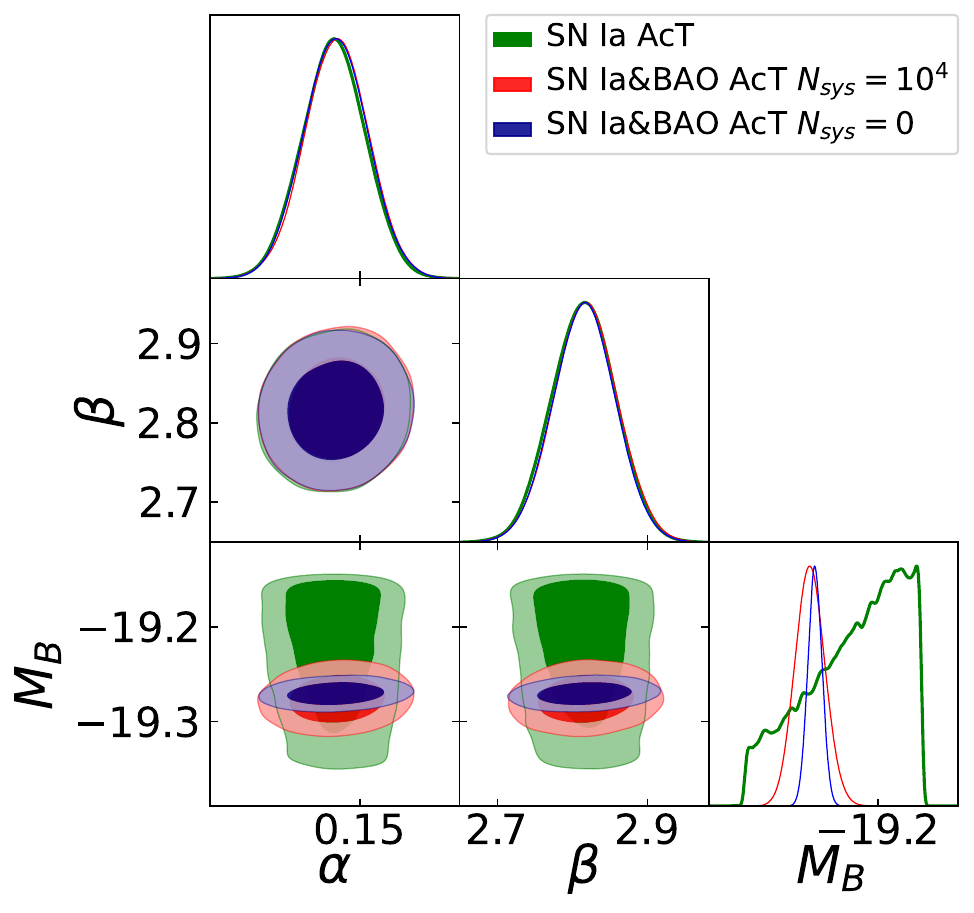}
	  \caption{\label{FigA.3}{\small ArcTanh }}
  \end{subfigure}%
  \caption{The contour maps and PDFs of the nuisance parameters $\alpha$, $\beta$ and $M_B$ in the SN Ia model for the three $f(R)$ model, constrained by the CSST SN Ia and SN Ia+BAO mock data with $N_{\text{sys}} = 0$ and $10^4\ h^{-3}{\rm Mpc^3}$.}
  \label{fig_NuisancePara}
\end{figure}

In Fig.\ref{fig_NuisancePara}, we show the posterior distribution of the SN Ia-related nuisance parameters under the Hu-Sawicki, Starobinsky, and ArcTanh $f(R)$ models after the MCMC process for a given prior. The details of the constraint results are shown in Table \ref{Table_Nuisance}.

\begin{table}[h]
    \centering
    \caption{The constraint results of the SNa-related nuisance parameters in $\Lambda \text{CDM}$, Hu-Sawicki, Starobinsky, and ArcTanh models restricted by the SN Ia and the SN Ia+BAO mock data with $N_{\text{sys}} = 0$ and $10^4\ h^{-3}{\rm Mpc^3}$.}
    \begin{tabular}{ccccc}
       \toprule
       Model&Dateset& $\alpha$ & $\beta$  & $M_B$  \\
       \midrule
       \multirow{3}{*}{$\Lambda\text{CDM}$} 
       &\tiny{SN Ia} &$0.147^{+0.004}_{-0.003}$&$2.818^{+0.040}_{-0.044}$ & $-19.213^{+0.034}_{-0.092}$\\
       &\tiny{SN Ia+BAO($N_{\mathrm{sys}} = 0$)} &$0.147^{+0.004}_{-0.003}$&$ 2.812^{+0.047}_{-0.037} $&$-19.155^{+0.016}_{-0.133}$  \\
       &\tiny{SN Ia+BAO($N_{\mathrm{sys}} = 10^4$)} &$ 0.147^{+0.003}_{-0.004} $&$  2.816^{+0.044}_{-0.039}$& $-19.281^{+0.019}_{-0.014}$  \\
       \midrule
       \multirow{3}{*}{$\text{HS}$} &\tiny{SN Ia} & $0.147^{+0.004}_{-0.003}$ & $2.816^{+0.041}_{-0.043}$ &  $-19.155^{+0.016}_{-0.133}$ \\ 
       &\tiny{SN Ia+BAO($N_{\mathrm{sys}}=0$) } & $0.148^{+0.003}_{-0.004}$ & $2.818^{+0.041}_{-0.044}$  &$-19.270^{+0.008}_{-0.008}$  \\
       &\tiny{SN Ia+BAO($N_{\mathrm{sys}}=10^4$) } & $0.147^{+0.004}_{-0.003}$ & $ 2.810^{+0.049}_{-0.034}$  &$-19.276^{+0.017}_{-0.016}$   \\
       \midrule 
       \multirow{3}{*}{$\text{St}$} &\tiny{SN Ia } & $0.147^{+0.003}_{-0.004} $ & $2.817^{+0.041}_{-0.043} $ & $-19.187^{+0.012}_{-0.107}$ \\
       &\tiny{SN Ia+BAO($N_{\mathrm{sys}}=0$) } & $0.147^{+0.004}_{-0.003}$ & $2.806^{+0.053}_{-0.031} $  & $-19.269^{+0.009}_{-0.010}$ \\
       &\tiny{SN Ia+BAO($N_{\mathrm{sys}}=10^4$) } & $0.148^{+0.003}_{-0.004} $ & $2.822^{+0.038}_{-0.046}  $ & $-19.274^{+0.015}_{-0.018}$  \\
       \midrule 
       \multirow{3}{*}{$\text{AcT}$} &\tiny{SN Ia } & $0.147^{+0.003}_{-0.004}$ & $ 2.805^{+0.052}_{-0.033} $ &  $-19.173^{+0.002}_{-0.115}$  \\
       &\tiny{SN Ia+BAO($N_{\mathrm{sys}}=0$) } & $0.147^{+0.003}_{-0.004}$ & $2.816^{+0.044}_{-0.039}$ & $-19.281^{+0.019}_{-0.014} $ \\
       &\tiny{SN Ia+BAO($N_{\mathrm{sys}}=10^4$) } & $0.148^{+0.003}_{-0.004}$ & $2.822^{+0.038}_{-0.046} $ & $-19.273^{+0.014}_{-0.019}$  \\
       \bottomrule  
    \end{tabular}
    \label{Table_Nuisance}
\end{table}

\newpage  
\bibliographystyle{raa}
\bibliography{bibtex}

\end{document}